\documentclass[aps,prb,floatfix,twocolumn,amsmath,amssymb]{revtex4}
\usepackage{graphicx}
\usepackage{dcolumn}
\usepackage{bm}
\usepackage{color}

\begin{document}

\title{Hidden Mott transition and large-$U$ superconductivity in the two-dimensional Hubbard model}
\author{Luca F. Tocchio, Federico Becca, and Sandro Sorella}
\affiliation{
 CNR-IOM-Democritos National Simulation Centre 
 and International School for Advanced Studies (SISSA), 
 Via Bonomea 265, I-34136, Trieste, Italy
            }

\date{\today} 

\begin{abstract}
We consider the one-band Hubbard model on the square lattice by using variational and Green's function Monte Carlo methods, where the 
variational states contain Jastrow and backflow correlations on top of an uncorrelated wave function that includes BCS pairing and magnetic
order. At half filling, where the ground state is antiferromagnetically ordered for any value of the on-site interaction $U$, we can identify 
a hidden critical point $U_{\rm Mott}$, above which a finite BCS pairing is stabilized in the wave function. The existence of this point is 
reminiscent of the Mott transition in the paramagnetic sector and determines a separation between a Slater insulator (at small values of $U$),
where magnetism induces a potential energy gain, and a Mott insulator (at large values of $U$), where magnetic correlations drive a kinetic 
energy gain. Most importantly, the existence of $U_{\rm Mott}$ has crucial consequences when doping the system: We observe a tendency to phase 
separation into a hole-rich and a hole-poor region only when doping the Slater insulator, while the system is uniform by doping the Mott 
insulator. Superconducting correlations are clearly observed above $U_{\rm Mott}$, leading to the characteristic dome structure in doping.
Furthermore, we show that the energy gain due to the presence of a finite BCS pairing above $U_{\rm Mott}$ shifts from the potential to the 
kinetic sector by increasing the value of the Coulomb repulsion.
\end{abstract}


\maketitle

\section{Introduction}\label{sec:intro}

The emergence of high-temperature superconductivity upon doping in the two-dimensional Copper-oxide planes of insulating antiferromagnetic 
cuprate materials is still a great puzzle in condensed matter physics, after many years from the first experimental evidence~\cite{bednorz1986}. 
In particular, from a theoretical point of view, a still open question is about the mechanism behind the appearance of the superconducting state.
One conservative approach is to explain the electron pairing by invoking the electron-phonon coupling, as in standard BCS theory; an alternative
approach is based upon the so-called resonating-valence bond (RVB) theory, as originally proposed by Anderson~\cite{anderson1987}, in which
superconductivity emerges from a Mott insulator that possesses preformed electron pairs.

In order to address the role of electron correlation on a lattice, one of the simplest models is the single-band Hubbard model, defined as:
\begin{equation}\label{eq:Hubbard}
{\cal H} = -t \sum_{\langle i,j\rangle,\sigma} c^{\dagger}_{i,\sigma}c^{\phantom{\dagger}}_{j,\sigma} + \textrm{h.c.} 
+U\sum_i n_{i,\uparrow}n_{i,\downarrow},
\end{equation}
where the hopping amplitude between nearest-neighbor sites on the square lattice and the on-site Coulomb repulsion are denoted by $t$ and $U$,
respectively; then, $c^{\dagger}_{i,\sigma} (c^{\phantom{\dagger}}_{i,\sigma})$ is the creation (annihilation) operator for an electron of 
spin $\sigma$ on site $i$ and $n_{i,\sigma}=c^{\dagger}_{i,\sigma}c^{\phantom{\dagger}}_{i,\sigma}$ is the electron density (per spin) on 
site $i$. Despite its simplicity, the Hubbard model has been proposed to capture the essential physics of high-temperature superconductivity 
and interaction-driven metal-insulator transitions~\cite{dagotto1994,imada1998}. The exact solution of this model is not available in spatial
dimensions greater than one for generic values of electron densities; instead, Monte Carlo methods provide numerically exact solutions at half 
filling~\cite{hirsch1985,white1989,leblanc2015,qin2016,vitali2016}, predicting an insulating ground state with antiferromagnetic order for
positive values of $U/t$. A summary of state-of-the-art numerical methods to address the Hubbard model at different interactions and dopings 
can be found in Ref.~\onlinecite{leblanc2015}.

When antiferromagnetism is suppressed, a metal-insulator transition can be identified at half filling for a critical value of the Coulomb 
repulsion~\cite{brinkman1970,georges1996}, separating a metallic state, for small values of the Coulomb repulsion, from a Mott insulator, for 
a larger Coulomb interaction. The non-magnetic sector of the Hubbard model has been the starting point of several studies to investigate the 
emergence of superconductivity upon doping, for instance with cluster extensions of dynamical mean-field theory (DMFT)~\cite{sordi2012,gull2013} 
or with variational Monte Carlo (VMC)~\cite{eichenberger2007,tocchio2012,yokoyama2013}. All these studies suggest that a value of the Coulomb repulsion of the 
order of the bandwidth is necessary for stabilizing superconductivity at finite doping, the symmetry of the order parameter being $d$-wave. 
Furthermore, the appearance of superconductivity by doping the Hubbard model has been studied also by the diagrammatic Monte 
Carlo method, indicating a BCS-type instability (with $d$-wave symmetry) for dopings smaller than $40\%$, at $U/t\le 4$~\cite{deng2015}. 
Analogous results have been obtained also by a weak-coupling renormalization-group study, even if in this latter case the presence of 
a next-nearest neighbor hopping is important to stabilize superconductivity at finite doping~\cite{eberlein2014}. In the line of weak-coupling 
approaches, a quantum critical point, hidden under the superconducting dome, has been proposed as a mechanism to generate the high-temperature 
superconductivity, induced by a pairing instability stronger than the BCS logarithmic divergence~\cite{yang2011}. The idea of a critical doping 
has been also postulated a few years ago, in connection with the formation of charge-density waves and their relation with the formation of 
superconducting pairs~\cite{castellani1995}. 

At small doping, strong antiferromagnetic correlations are present, possibly leading to a region where superconductivity and long-range 
antiferromagnetic order coexist~\cite{assaad1998,lichtenstein2000,reiss2007,kancharla2008,markiewicz2010,sato2016}. Moreover, when 
antiferromagnetism is taken into account, phase separation may also occur, i.e., a non-uniform charge distribution, showing distinct hole-rich 
and hole-poor regions, as originally proposed in Ref.~\onlinecite{emery1990}. The presence of long-range Coulomb interactions in realistic 
materials would then ``frustrate'' phase separation, eventually leading to charge density states or stripes. We remark that the formation of 
striped phases is not necessarily related to the effect of long-range Coulomb interactions, since they can just be driven by a competition 
between kinetic and super-exchange energies~\cite{white1998}. The appearance of phase separation and the formation of stripes have been deeply 
discussed in the $t{-}J$ model, which describes the strong-coupling limit of the Hubbard model. Here, the tendency to phase separation has been 
questioned  by both density-matrix renormalization group (DMRG)~\cite{white2000}, which suggested the presence of stripes, and Green's Function 
Monte Carlo (GFMC) with the fixed-node (FN) approximation~\cite{lugas2006}, which instead supported an homogeneous and superconducting ground 
state. After many years of investigations, the presence of stripes in the ground state is still an open question~\cite{corboz2014}.

In the Hubbard model, a clear tendency of phase separation has been recently pointed out for $U/t \lesssim 4$, by using the auxiliary-field 
quantum Monte Carlo (AFQMC) method with modified boundary conditions~\cite{sorella2015}; for larger values of $U/t$, some indications have 
been provided by the variational cluster approach (VCA)~\cite{aichhorn2007}, by AFQMC with constrained path~\cite{chang2008}, and by 
VMC~\cite{tocchio2013,misawa2014}. However, determining the presence of phase separation in the Hubbard model is a difficult task, due to its 
strong dependence on the accuracy of the states that are used to compute the energy. In particular, it has been shown that phase separation is 
more favorable for less accurate variational states~\cite{becca2000}. The formation of stripes in the Hubbard model, possibly favored with 
respect to a homogeneous superconducting state, has been also considered~\cite{corboz2016}. In this respect, a recent density matrix embedding 
theory (DMET) study, performed up to $U/t=8$, did not show clear evidences for a striped ground state~\cite{zheng2016}.

In this paper, we perform a systematic study of the Hubbard model on the square lattice, by using accurate variational wave functions, which 
include both superconductivity and magnetism. At half filling, our VMC results indicate that, inside the antiferromagnetic phase, there is
a hidden transition at a finite value of the Coulomb repulsion $U_{\rm Mott}$, above which a finite BCS pairing is stabilized by energy 
minimization (in addition to magnetic order that is present for any finite value of the interaction strength $U$). We relate $U_{\rm Mott}$ 
with a crossover separating a Slater insulator at low values of $U/t$, where magnetism induces a potential energy gain, from a Mott insulator 
at large values of $U/t$, where magnetic correlations drive a kinetic energy gain. More importantly, the existence of a ``critical'' value of 
the Coulomb repulsion at half filling has crucial consequences on the behavior of the model at finite doping. First, we consider the issue of 
phase separation. While at the level of VMC, the model is prone to phase separation for all values of $U/t$, the more accurate energies provided 
by GFMC, within the FN approximation, allow us to clearly distinguish two regimes: When doping a Slater insulator, phase separation is obtained; 
instead, a homogeneous density distribution is preferred when doping the Mott insulator. Most importantly, we observe finite long-range pairing 
correlations, with weak finite-size effects and a characteristic dome structure, only when $U>U_{\rm Mott}$; for $U<U_{\rm Mott}$, there 
are strong size effects in the pairing correlations, which may survive only in a small part of the phase diagram. Finally, we analyze the 
behavior of the condensation energy, i.e., the energy gain due to the presence of a finite BCS pairing. We find that it changes from being 
potential driven to kinetic driven by increasing the value of the Coulomb repulsion and its maximum value is always located at the doping 
where magnetic correlations in the wave function disappear. From our overall results, we surmise that interaction is the crucial mechanism 
to observe sizable superconducting pairing, with a hidden critical value of the Coulomb repulsion that may be observed already within the 
antiferromagnetic phase at half filling.

The paper is organized as follows: In Sec.~\ref{sec:methods}, we present the variational wave functions that are used in this work, as well as 
the Monte Carlo approaches; in Sec.~\ref{sec:results}, we show the numerical results; finally in Sec.~\ref{sec:conc}, we draw our conclusions.
 
\begin{figure}
\includegraphics[width=0.8\columnwidth]{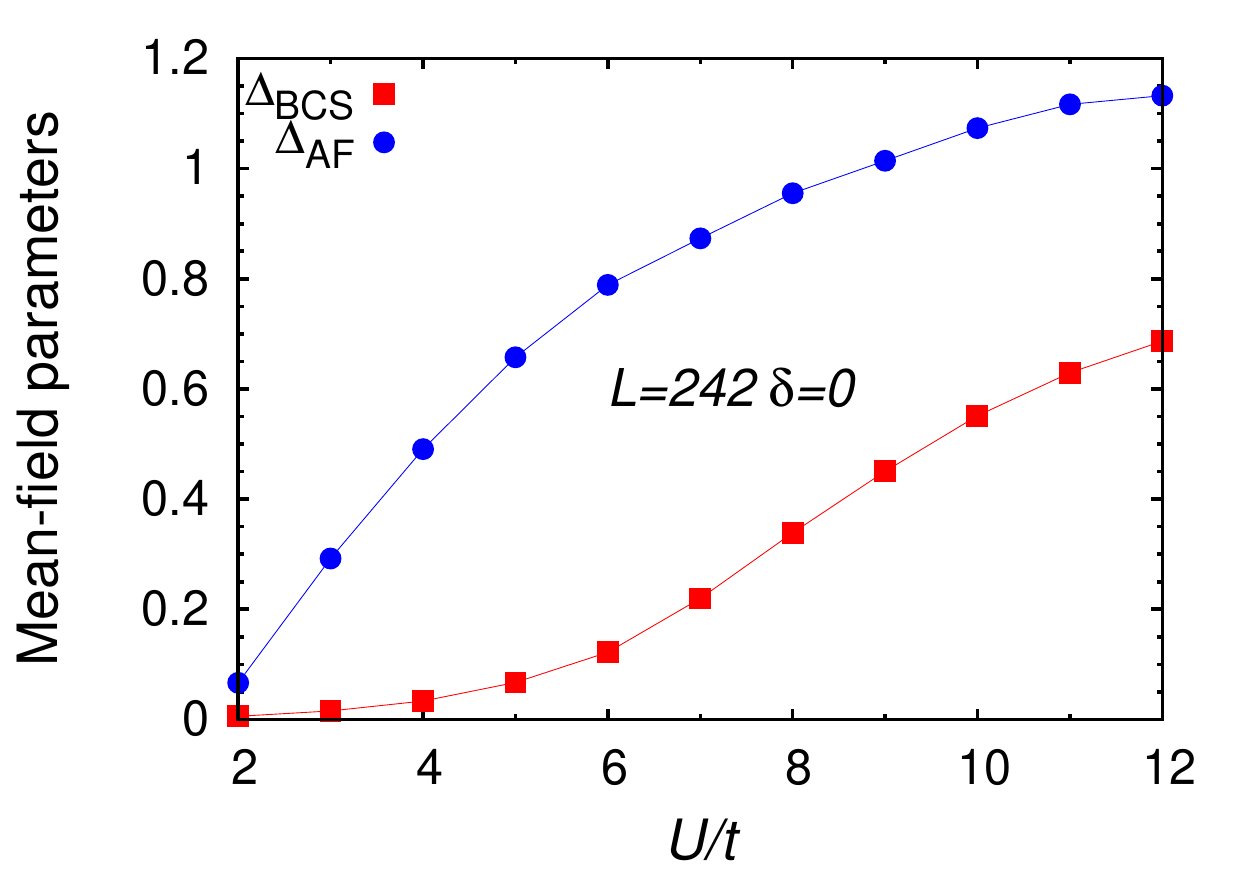}
\caption{\label{fig:HF_canonical}
(Color online) Mean-field variational parameters $\Delta_{\textrm{AF}}$ (blue dots) and $\Delta_{\textrm{BCS}}$ (red squares) as a function of 
$U/t$ for the half-filled case on a $L=242$ lattice size.}
\end{figure}

\section{Variational and Green's function Monte Carlo}\label{sec:methods}

Our numerical results are based on the definition of variational wave functions that approximate the ground-state properties beyond perturbative 
approaches. In order to compute expectation values over these correlated variational states a Monte Carlo sampling is necessary. The general form 
of our variational states is given by the Jastrow-Slater wave function that extends the original formulation by 
Gutzwiller~\cite{gutzwiller1963,yokoyama1987}:
\begin{equation}\label{eq:wavefunction}
|\Psi\rangle= {\cal P}_{N} {\cal P}_{S_z=0} {\cal J}|\Phi_0\rangle,  
\end{equation}
where $|\Phi_0\rangle$ is an uncorrelated state that corresponds to the ground state of the following uncorrelated Hamiltonian~\cite{gros1988,zhang1988}:
\begin{equation}\label{eq:meanfield}
\begin{split}
{\cal H}_{\rm{MF}} &= \sum_{k\sigma} \xi_k c^{\dagger}_{k\sigma} c^{\phantom{\dagger}}_{k\sigma}
+ \sum_{k} \Delta_k c^{\dagger}_{k\uparrow} c^{\dagger}_{-k\downarrow} + \rm{h.c.} \\
 &+ \Delta_{\textrm{AF}} \sum_i (-1)^{R_i} ( c^{\dagger}_{i\uparrow} c^{\phantom{\dagger}}_{i\uparrow} -
c^{\dagger}_{i\downarrow} c^{\phantom{\dagger}}_{i\downarrow} ),
\end{split}
\end{equation}
which includes a free-band dispersion:
\begin{equation}\label{eq:hopping}
\xi_k= -2 t(\cos k_x +\cos k_y) -\mu,
\end{equation}
a BCS pairing with $d$-wave symmetry:
\begin{equation}\label{eq:BCS}
\Delta_k= 2\Delta_{\textrm{BCS}}(\cos k_x -\cos k_y),
\end{equation}
as well as an antiferromagnetic term with N\'eel order. The parameters $\Delta_{\textrm{AF}}$, $\Delta_{\textrm{BCS}}$ and $\mu$ are optimized 
to minimize the variational energy (while $t=1$ sets the energy scale of the uncorrelated Hamiltonian). The effects of correlations are introduced 
by means of the so-called Jastrow factor ${\cal J}$~\cite{capello2005,capello2006}:
\begin{equation}\label{eq:jastrow}
{\cal J} = \exp \left ( -\frac{1}{2} \sum_{i,j} v_{i,j} n_{i} n_{j} \right ),
\end{equation}
where $n_{i}= \sum_{\sigma} n_{i,\sigma}$ is the electron density on site $i$ and $v_{i,j}$ (that include also the local Gutzwiller term for 
$i=j$) are pseudopotentials that are optimized for every independent distance $|{\bf R}_i-{\bf R}_j|$. Finally, ${\cal P}_{N}$ is a projector 
on the fixed number of particles $N$ and ${\cal P}_{S_z=0}$ is a projector onto the subspace with $S_z=0$.

A size-consistent and efficient way to further improve the correlated state $|\Psi\rangle$ for large on-site interactions is based on backflow 
correlations. In this approach, each orbital that defines the unprojected state $|\Phi_0\rangle$ is taken to depend upon the many-body 
configuration, such to incorporate virtual hopping processes~\cite{tocchio2008,tocchio2011}. All results presented here are obtained by fully 
incorporating the backflow corrections and optimizing individually every variational parameter in ${\cal H}_{\textrm{MF}}$, in the Jastrow 
factor ${\cal J}$, as well as for the backflow corrections~\cite{yunoki2006}. 

In general, on finite sizes, the presence of quantum-number projectors in the variational state may induce an energy gain, which however is
expected to vanish in the thermodynamic limit, see for instance Refs.~\onlinecite{tahara2008,rodriguez2012}. In our case, the projector onto
the subspace with exactly $N$ particles could introduce a spurious stabilization of a small BCS pairing, especially for small values of $U/t$.
Therefore, in order to reduce the finite-size effects, we also perform simulations without including ${\cal P}_{N}$ in the definition of the
wave function~(\ref{eq:wavefunction}), i.e., in the grand-canonical ensemble. In practice, this kind of simulation is performed by including 
in the Metropolis algorithm the option of creating or destroying pairs of electrons with opposite spin. The average number of particles is then
fixed via the inclusion of a chemical potential in the Hubbard Hamiltonian, namely ${\cal H} \to {\cal H}-\mu N$ (at half filling, $\mu=U/2$). 

The accuracy of the described variational states can be further increased by means of the GFMC method~\cite{trivedi1990}, based on the FN 
approximation~\cite{tenhaaf1995}. This approach allows us to systematically improve the variational results, still providing an upper bound 
to the exact ground-state energy. In practice, the best variational wave function $|\Psi_{\textrm{FN}}\rangle$, with the nodes constrained 
by the optimal variational state $|\Psi\rangle$, is extracted from an imaginary-time projection. A detailed description of the FN 
approximation can be found in Ref.~\onlinecite{lugas2006}, while a comparison of the accuracy of the method with other established numerical 
techniques is presented in Ref.~\onlinecite{leblanc2015}.

The accurate estimates of the FN energy can be used to evaluate the tendency of the system toward phase separation into undoped (with 
antiferromagnetic order) and hole-doped regions. Phase separation occurs when the stability condition $\partial^2 E(n)/\partial n^2 >0$ is 
violated, i.e., when the ground-state energy $E(n)$, as a function of electronic density $n$ ($n=N/L$, $L$ being the lattice size), is no 
longer convex. As introduced by Emery and collaborators~\cite{emery1990}, phase separation between a hole-rich phase and an antiferromagnetic 
one can be studied by looking at the energy per hole $\epsilon(\delta)$, defined as:
\begin{equation}\label{eq:Emery}
\epsilon(\delta)=\frac{E(\delta)-E(0)}{\delta}, 
\end{equation}
where $\delta=1-n$ is the hole density. In the thermodynamic limit, $\epsilon(\delta)$ is monotonically increasing in a stable phase, while it 
remains constant in presence of phase separation. On a finite system, the energy per hole has a minimum at a critical doping $\delta_c$, the 
system being unstable to phase separation for $\delta<\delta_c$. 

Finally, the calculation of expectation values of non-local operators (like for example pairing-pairing correlations) $O$ in the FN method can 
be done by using the so-called mixed-average correction~\cite{foulkes2001}. First, we need to compute the mixed average, that is a biased 
estimator of the quantum average:
\begin{equation}
\langle O\rangle_{\textrm{MA}} = \frac{\langle \Psi|O|\Psi_{\textrm{FN}}\rangle}{\langle \Psi|\Psi_{\textrm{FN}}\rangle}, 
\end{equation}
then, the true expectation value can be approximated with the formula:
\begin{equation}\label{eq:Ceperleycorr}
\frac{\langle \Psi_{\textrm{FN}}|O|\Psi_{\textrm{FN}}\rangle}{\langle \Psi_{\textrm{FN}}|\Psi_{\textrm{FN}}\rangle} \approx
2\langle O \rangle_{\textrm{MA}} - \langle O\rangle_{\textrm{VMC}},
\end{equation}
where
\begin{equation}
\langle O\rangle_{\textrm{VMC}}=\frac{\langle \Psi|O|\Psi\rangle}{\langle \Psi|\Psi\rangle}
\end{equation}
indicates the variational estimate of the expectation value of the operator $O$ over the wave function $|\Psi\rangle$. This approach is justified 
provided the variational wave function $|\Psi\rangle$ is close to the FN state $|\Psi_{\textrm{FN}}\rangle$ and is expected to hold in our case, 
given the good quality of a variational state, where the effect of correlations is incorporated both in the amplitudes (via the Jastrow factor) 
and in the signs (via the backflow corrections).

All the simulations are performed on 45-degree tilted clusters with $L=2l^2$ sites, $l$ being an odd integer.

\begin{figure*}[t]
\includegraphics[width=1.5\columnwidth]{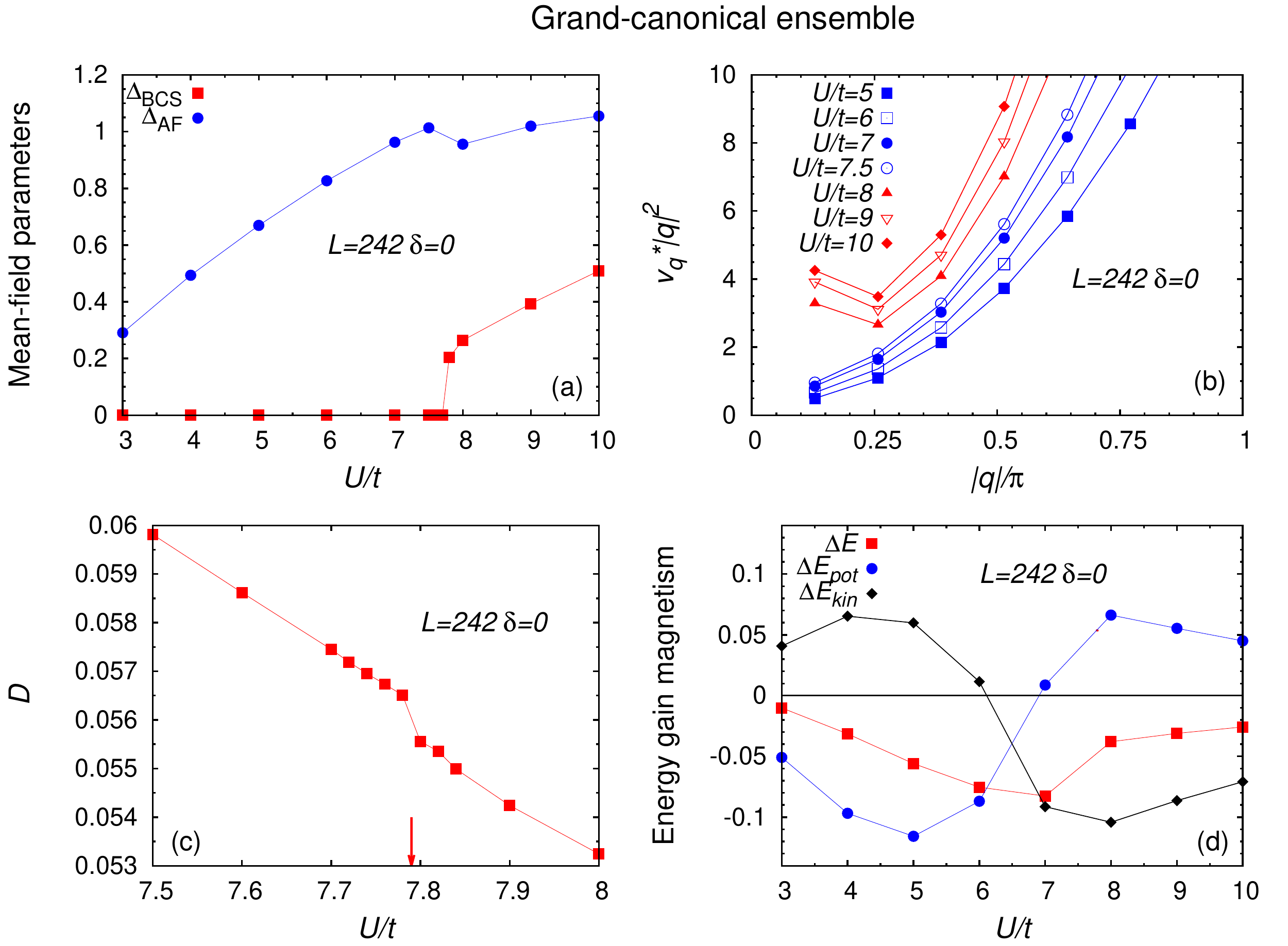}
\caption{\label{fig:BCS-hf}
(Color online) All the simulations shown are performed in the grand-canonical ensemble, at half filling on a $L=242$ lattice size.  
(a): Mean-field variational parameters $\Delta_{\textrm{AF}}$ (blue dots) and $\Delta_{\textrm{BCS}}$ (red squares) as a function 
of $U/t$. (b): Fourier transform of the Jastrow factor $v(q)$, multiplied by $|q|^2$, as a function of $|q|/\pi$, for various 
values of $U/t$. The $q$ points are taken on the path in the Brillouin zone connecting $\Gamma=(0,0)$ to $M=(\pi,\pi)$. (c): 
Density of double occupancies $D$ as a function of $U/t$. The arrow indicates the location of $U_{\rm Mott}$. (d): Energy gain 
$\Delta E=E_{\textrm{magn}}-E_{\textrm{nomagn}}$ (red squares), due to the presence of magnetism in the wave function; the potential 
$\Delta E_{\textrm{pot}}=E^{\textrm{pot}}_{\textrm{magn}}-E^{\textrm{pot}}_{\textrm{nomagn}}$ (blue dots) and kinetic 
$\Delta E_{\textrm{kin}}=E^{\textrm{kin}}_{\textrm{magn}}-E^{\textrm{kin}}_{\textrm{nomagn}}$ (black diamonds) contributions are also 
shown.}
\end{figure*}

\section{Results}\label{sec:results}

\subsection{The half-filled case}

Let us start by recalling the previous VMC results for the Hubbard model on the square lattice, when focusing on the non-magnetic sector, 
i.e., without the inclusion of magnetism in the variational state. In this case, a transition driven solely by electronic correlation, 
the so-called Mott transition, occurs at a critical value of the Coulomb repulsion $U_c/t$, that corresponds to $U_c/t\sim 7.5\pm 0.5$, 
when backflow correlations are not included~\cite{tocchio2016} and to $U_c/t\sim 5.5 \pm 0.5$, in the presence of backflow 
corrections~\cite{tocchio2011}. Since backflow corrections favor the recombination of holon-doublon couples into single occupied sites, 
this leads to an improvement in the description of the insulating phase, which can be stabilized at lower values of $U/t$~\cite{tocchio2011}. 
We remark that, even if the Jastrow factor is the driving force for the system to be an insulator, the Mott state is also characterized by a 
finite BCS pairing among the electrons, reproducing the RVB state, originally proposed by Anderson~\cite{anderson1987}.   

When magnetic order is allowed, the nesting properties of the Fermi surface drive the system to be an insulator with long-range N\'eel 
order for any $U>0$. However, while the antiferromagnetic coupling $\Delta_{\textrm{AF}}$ is always finite, our VMC results with fixed 
number of electrons, namely when using the wave function~(\ref{eq:wavefunction}), suggest that a reminiscence of the Mott transition can 
still be seen in the variational parameters, with a smooth crossover between a small-$U$ region with no relevant BCS pairing and a large-$U$ 
region with finite $\Delta_{\textrm{BCS}}$, see Fig.~\ref{fig:HF_canonical}. However, non-negligible size effects are present, especially 
at small values of $U/t$, preventing us to perform a clear size scaling of the variational parameters. Therefore, in order to reduce 
finite-size effects, we have performed the simulations within the grand-canonical ensemble, as described in Sec.~\ref{sec:methods}. 
These results are sharpened, with the clear identification of a hidden critical point located at $U_{\rm Mott}/t\simeq 7.8$ above which 
the BCS pairing becomes finite, see Fig.~\ref{fig:BCS-hf}(a). We remark that the value of $U_{\rm Mott}$ is not significantly affected 
by the lattice size. Indeed, it falls in the interval $7.5 < U_{\rm Mott}/t <8$ for lattice sizes ranging from $L=162$ to $L=338$. 

While the insulating nature of the model at half filling is guaranteed by the presence of a finite antiferromagnetic field $\Delta_{\textrm{AF}}$ 
that opens a gap already at the uncorrelated level, a reminiscence of the Mott transition can be seen also in the behavior of the Jastrow factor. 
Indeed, as shown in Fig.~\ref{fig:BCS-hf}(b), the Fourier transform of the Jastrow factor $v_q$ changes its small-$q$ behavior from $1/q$ to 
$1/q^2$ at $U_{\rm Mott}$, as it would do at the true Mott transition~\cite{capello2005,capello2006,yokoyama2011}, where the Jastrow factor 
embodies a crucial long-range attraction between doubly occupied and empty sites, keeping them bounded in the Mott phase. A Jastrow factor 
proportional to $1/q^2$ is also able to suppress the superconducting long-range order implied by the BCS pairing of $|\Phi_0\rangle$. 
More importantly, a sudden change in the average density of double occupancies
\begin{equation}
D=\left\langle \frac{1}{L} \sum_i n_{i,\uparrow} n_{i,\downarrow} \right\rangle_{\textrm{VMC}}
\end{equation}
is observed at the ``critical'' point where $\Delta_{\textrm{BCS}}$ becomes finite, as shown in Fig.~\ref{fig:BCS-hf}(c). This feature suggests 
that the appearance of the finite BCS pairing in the variational state coincides with a change in the nature of the magnetic insulator at half 
filling. This change can be understood by investigating the role of magnetism on the variational energy. At small $U/t$, the ground state exhibits 
the so-called Slater magnetism in which the presence of a finite magnetic term suppresses double occupancies and consequently induces a potential 
energy gain. On the contrary, for large $U/t$, magnetism is favored by the super-exchange coupling $J=4t^2/U$ and consequently drives a kinetic 
energy gain, leading to the so-called Mott magnetism. The crossover between these two regimes has been investigated by many authors in the past: Even if 
it has been proposed that the Slater mechanism is absent in the Hubbard model~\cite{moukouri2001}, successive works show a general consensus 
on the existence of a sharp crossover between the two regimes, with the precise location of it depending on the method 
of investigation~\cite{kyung2003,korbel2003,pruschke2003,plekhanov2005,gull2008,taranto2012}. 

The energy gain that is due to magnetism goes to zero for $U\to 0$ and for $U \to \infty$ and is expected to have a maximum when passing from 
Slater to Mott magnetism. Indeed, our results show the existence of a value of the Coulomb repulsion where the total energy gain due to magnetism 
$\Delta E=E_{\textrm{magn}}-E_{\textrm{nomagn}}$ is maximal and, at the same time, the contributions coming from potential and kinetic energies 
change signs, see Fig.~\ref{fig:BCS-hf}(d). While $E_{\textrm{magn}}$ is given by the full wave function, as defined in Sec.~\ref{sec:methods}, 
in the computation of $E_{\textrm{nomagn}}$ we just set $\Delta_{\textrm{AF}}=0$ in the uncorrelated Hamiltonian of Eq.~(\ref{eq:meanfield}). 
The results shown in Fig.~\ref{fig:BCS-hf}(d) indicate that the appearance of a finite BCS pairing in the wave function at $U_{\rm Mott}$ affects 
the magnetic properties of the model, inducing a clear change between Slater and Mott type of magnetism. This fact has important consequences 
on the behavior of the model as a function of doping, as presented below.

\begin{figure}
\includegraphics[width=0.8\columnwidth]{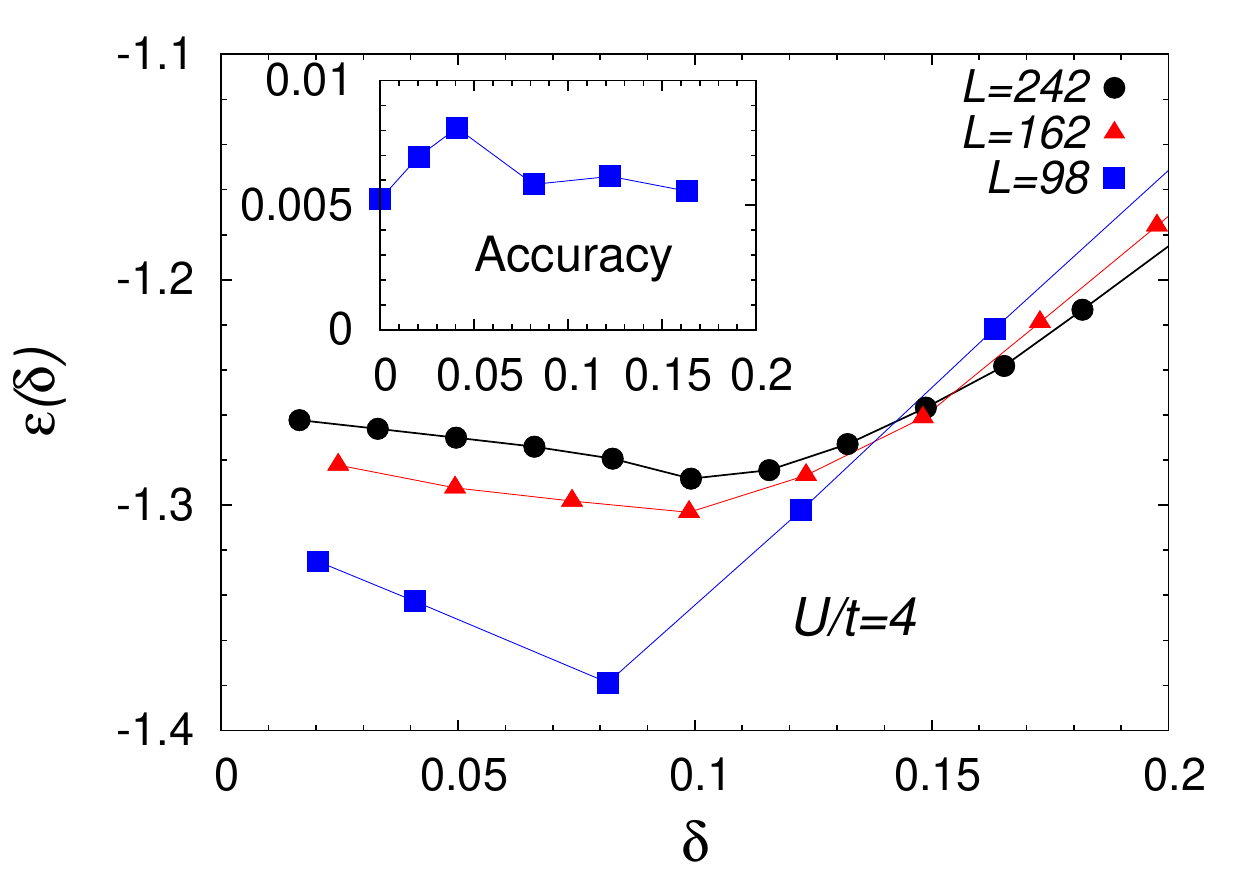}
\caption{\label{fig:PSU4}
(Color online) Energy per hole $\epsilon(\delta)$ for the FN energies at $U/t=4$, as a function of the hole doping $\delta$, for three 
lattice sizes: $L=98$ (blue squares), $162$ (red triangles) and $242$ (black circles). The error bars are smaller than the symbol size. 
The inset shows the accuracy of the FN energies with respect to the linearized AFQMC ones~\cite{sorella2011}, defined as 
$(E_{\textrm{AFQMC}}-E_{\textrm{FN}})/E_{\textrm{AFQMC}}$ for $U/t=4$, as a function of hole doping on the $L=98$ lattice size.}
\end{figure}

\begin{figure}
\includegraphics[width=0.8\columnwidth]{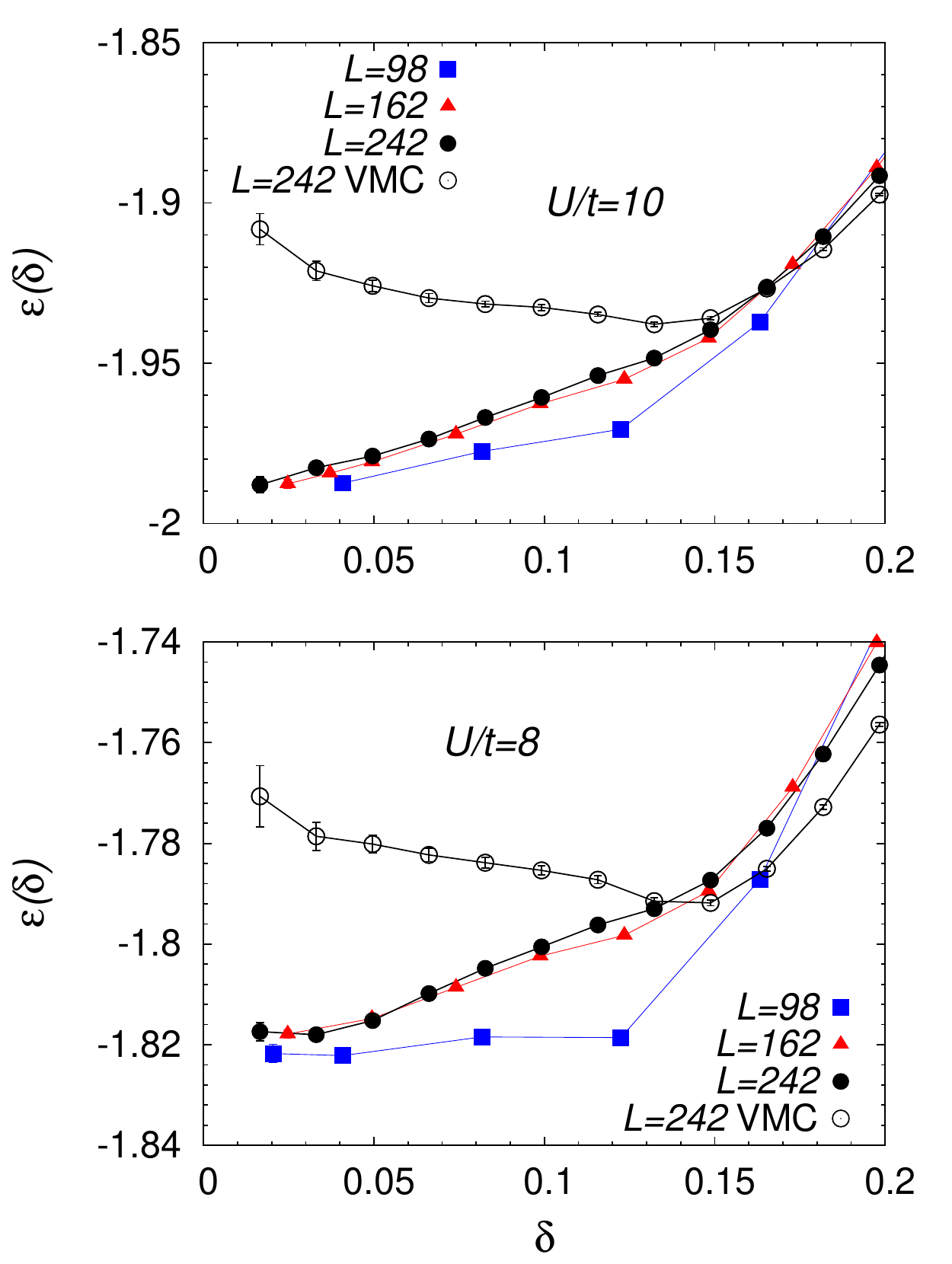}
\caption{\label{fig:PSU8U10}
(Color online) Lower panel: Energy per hole $\epsilon(\delta)$ for the FN energies at $U/t=8$, as a function of the hole doping $\delta$, 
for three lattice sizes: $L=98$ (blue squares), $162$ (red triangles) and $242$ (black circles). The VMC results are also shown on the $L=242$ 
lattice size (empty black circles). Upper panel: The same as the lower panel for $U/t=10$.}
\end{figure}

\begin{table*}
\caption{\label{table:Energies}
VMC and FN energies as a function of the number of holes $n_h$ at $U/t=4$ and $U/t=10$ on the $L=242$ lattice size. The number in brackets 
denotes the error on the last digit.}
\begin{tabular}{ccccc}
\hline
$n_h$ & $E/t$ (VMC $U/t=4$) & $E/t$ (FN $U/t=4$) & $E/t$ (VMC $U/t=10$) & $E/t$ (FN $U/t=10$) \\
\hline\hline
 0        &  -0.85496(2)    &  -0.85725(3)      &  -0.42712(6)     &  -0.43206(2)     \\
 8        &  -0.89642(2)    &  -0.89910(2)      &  -0.49064(4)     &  -0.49760(2)     \\
 16       &  -0.93888(2)    &  -0.94148(1)      &  -0.55471(3)     &  -0.56255(2)     \\
 24       &  -0.98315(1)    &  -0.98501(1)      &  -0.61879(3)     &  -0.62651(2)     \\       
 32       &  -1.02407(1)    &  -1.02557(1)      &  -0.68337(3)     &  -0.68970(2)     \\
 40       &  -1.06057(1)    &  -1.06192(1)      &  -0.74559(2)     &  -0.75045(1)     \\      
 48       &  -1.09173(1)    &  -1.09285(1)      &  -0.80348(2)     &  -0.80724(1)     \\
 56       &  -1.11919(1)    &  -1.12027(1)      &  -0.85625(2)     &  -0.85938(1)     \\
\hline \hline 
\end{tabular}
\end{table*}

\subsection{Phase separation}

Here, we consider the tendency to phase separation as a function of the interaction strength $U$. Our results show that $U_{\rm Mott}$, where 
a finite BCS pairing in the uncorrelated state starts to develop at half filling, separates two different regimes also at finite doping. 
For $U \lesssim U_{\rm Mott}$, the variational state contains only a magnetic order parameter $\Delta_{\textrm{AF}}$ and phase separation 
arises upon doping; by contrast, for $U \gtrsim U_{\rm Mott}$, the presence of a finite BCS pairing $\Delta_{\textrm{BCS}}$ inhibits phase 
separation, leading to a superconducting state at finite hole dopings. In order to determine the existence of phase separation we use the 
energy per hole of Eq.~(\ref{eq:Emery}) for the VMC and FN energies.

As already discussed in Ref.~\onlinecite{becca2000}, the evaluation of phase separation is strongly affected by the accuracy of the states 
that are used to compute the energy, phase separation being more favorable for less accurate variational wave functions. In fact, at the pure 
VMC level, we find that phase separation dominates the low-doping regime of the phase diagram for a wide range of interaction strengths $U$, 
as shown below. The main problem is that a slight difference in the accuracy for different dopings induces huge errors in the energy per hole, 
especially close to half filling, where $\delta$ is small. In this case, the application of the grand-canonical approach does not help to 
stabilize a uniform phase and phase separation still appears for all values of $U/t$. This result is in the line of an independent VMC 
calculation~\cite{misawa2014}, which related the onset of superconductivity with the proximity to phase separation. 

Therefore, we move to the FN results, which give a much more accurate energy estimate when varying the hole doping. Unfortunately, the 
grand-canonical approach cannot be used within the GFMC method, since the imaginary-time projection is driven by the Hubbard Hamiltonian, 
which conserves the number of particles. At $U/t=4$, our results show that a rather wide region of phase separation occurs up to $\delta\sim 0.1$, 
see Fig.~\ref{fig:PSU4}. This result is in agreement with recent estimates provided by the AFQMC method with modified boundary 
conditions~\cite{sorella2015}. Moreover, the accuracy of our energies with respect to the AFQMC results on the same lattice size is remarkably 
good, being always lower than $10^{-2}$, as shown in the inset of Fig.~\ref{fig:PSU4}. The situation changes drastically when the value of 
$U$ increases above the threshold set by the appearance of a finite BCS pairing at half filling. In Fig.~\ref{fig:PSU8U10}, we present the 
energy per hole at $U/t=8$, i.e., just above $U_{\rm Mott}$. While the results at $L=98$ show that there is some tendency towards phase 
separation, by increasing the lattice size the appearance of phase separation becomes confined to a small doping interval $\delta \lesssim 0.04$, 
for both $L=162$ and $L=242$. The fact that the curves obtained on these two lattice sizes are almost superimposed suggests that we are close 
to the thermodynamic limit already on the $L=162$ lattice. Then, by increasing the Coulomb repulsion up to $U/t=10$, phase separation does not 
occur for the values of doping that can be studied with the available clusters (i.e., $\delta \gtrsim 0.02$), see Fig.~\ref{fig:PSU8U10}. 
Also in this case, some finite size effects are still visible on the $L=98$ lattice size, while the curves at $L=162$ and $L=242$ are almost 
coincident. 

The absence of phase separation for large values of the Coulomb repulsion is in marked disagreement with VMC results: Indeed, as shown in 
Fig.~\ref{fig:PSU8U10}, within VMC calculations, phase separation appears also at $U/t=10$ and at $U/t=8$, while it is absent once we improve
the accuracy of the calculations. The VMC and FN energies for several values of doping at $U/t=4$ and $U/t=10$ are reported in 
Table~\ref{table:Energies}.

\begin{figure*}
\includegraphics[width=1.5\columnwidth]{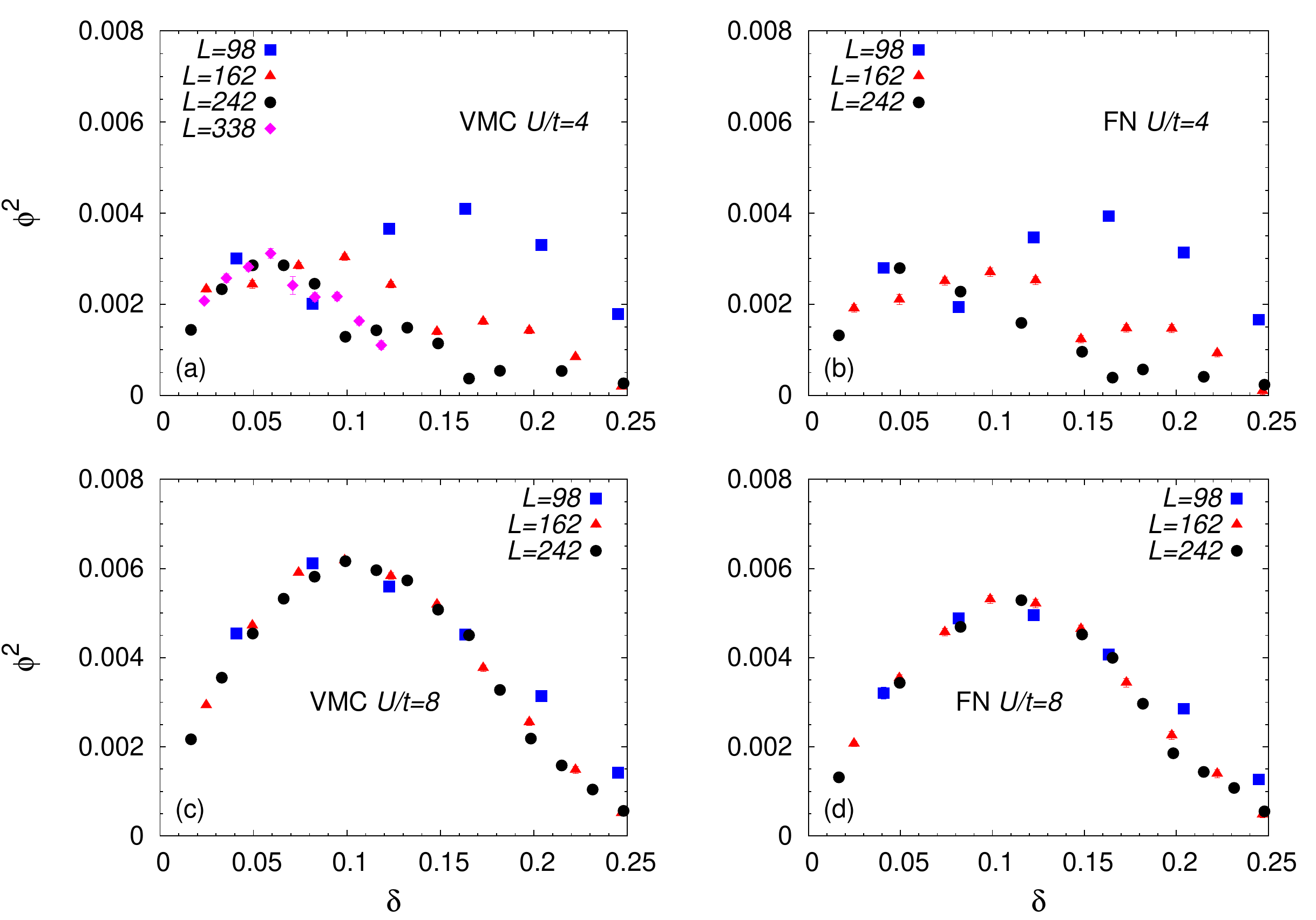}
\caption{\label{fig:phi}
(Color online) Panel (a): Superconducting order parameter squared $\phi^2$ of Eq.~(\ref{eq:phi}) as a function of doping $\delta$ for $U/t=4$, 
computed over the optimal variational state. Data are shown for $L=98$ (blue squares), $162$ (red triangles), $242$ (black circles), $338$ 
(purple diamonds). Panel (b): The same as in panel (a) computed within the FN approximation, with the mixed-average correction of 
Eq.~(\ref{eq:Ceperleycorr}). Data are shown for $L=98$ (blue squares), $162$ (red triangles), $242$ (black circles). Panel (c): The same as 
panel (a), but for $U/t=8$. Panel (d): The same as panel (b), but for $U/t=8$.} 
\end{figure*}

\subsection{Superconducting properties}

Here, we show the results of the pairing-pairing correlations:
\begin{equation}
\langle \Delta(r)\rangle=\langle S^{\phantom{\dagger}}_{r}S^{\dagger}_0 \rangle,
\end{equation}
where $S^{\dagger}_r= c^{\dagger}_{r\uparrow}c^{\dagger}_{r+x\downarrow} -c^{\dagger}_{r\downarrow}c^{\dagger}_{r+x\uparrow}$, can be easily 
evaluated in the VMC approach. The long-distance limit of the correlations gives an estimate of the superconducting order parameter:
\begin{equation}\label{eq:phi}
\phi^2=\lim_{r \to \infty}\Delta(r). 
\end{equation}
In analogy to what has been done in previous studies, for the Hubbard and $t{-}J$ models~\cite{tocchio2012,paramekanti2001,spanu2008}, we 
report the pairing-pairing correlations at the largest distance for different lattice sizes, to infer the behavior of $\phi^2$ in the 
thermodynamic limit. The VMC results, obtained by considering the best variational state, are shown in Fig.~\ref{fig:phi} for $U/t=4$ and $8$. 
They clearly show that, for $U \gtrsim U_{\rm Mott}$ (i.e., for $U/t=8$), a finite superconducting order parameter in the thermodynamic limit 
can be obtained upon doping, with a characteristic dome structure. Assuming that the critical temperature scales with $\phi$, our results 
locate the optimal doping at $\delta \approx 0.1$. A similar behavior is obtained also at $U/t=10$ (not shown). 

On the contrary, by doping the Slater insulator at $U/t=4$, the superconducting order parameter suffers from strong finite size effects for 
$\delta \gtrsim 0.1$, indicating that no sizable superconductivity survives in the thermodynamic limit. Nevertheless, small superconducting 
correlations might be finite for $\delta \lesssim 0.1$; however, we must stress that, in this region, the system shows a tendency to phase 
separation, see Fig.~\ref{fig:PSU4}, so that superconductivity is hindered by the non-homogeneous spatial distribution of electrons.
In order to assess the validity of these VMC results, we also consider FN estimates. In particular, we compute $\phi^2$ with the mixed-average 
correction of Eq.~(\ref{eq:Ceperleycorr}). The results confirm the above described behavior for both regimes, see Fig.~\ref{fig:phi}.

\begin{figure*}
\includegraphics[width=1.5\columnwidth]{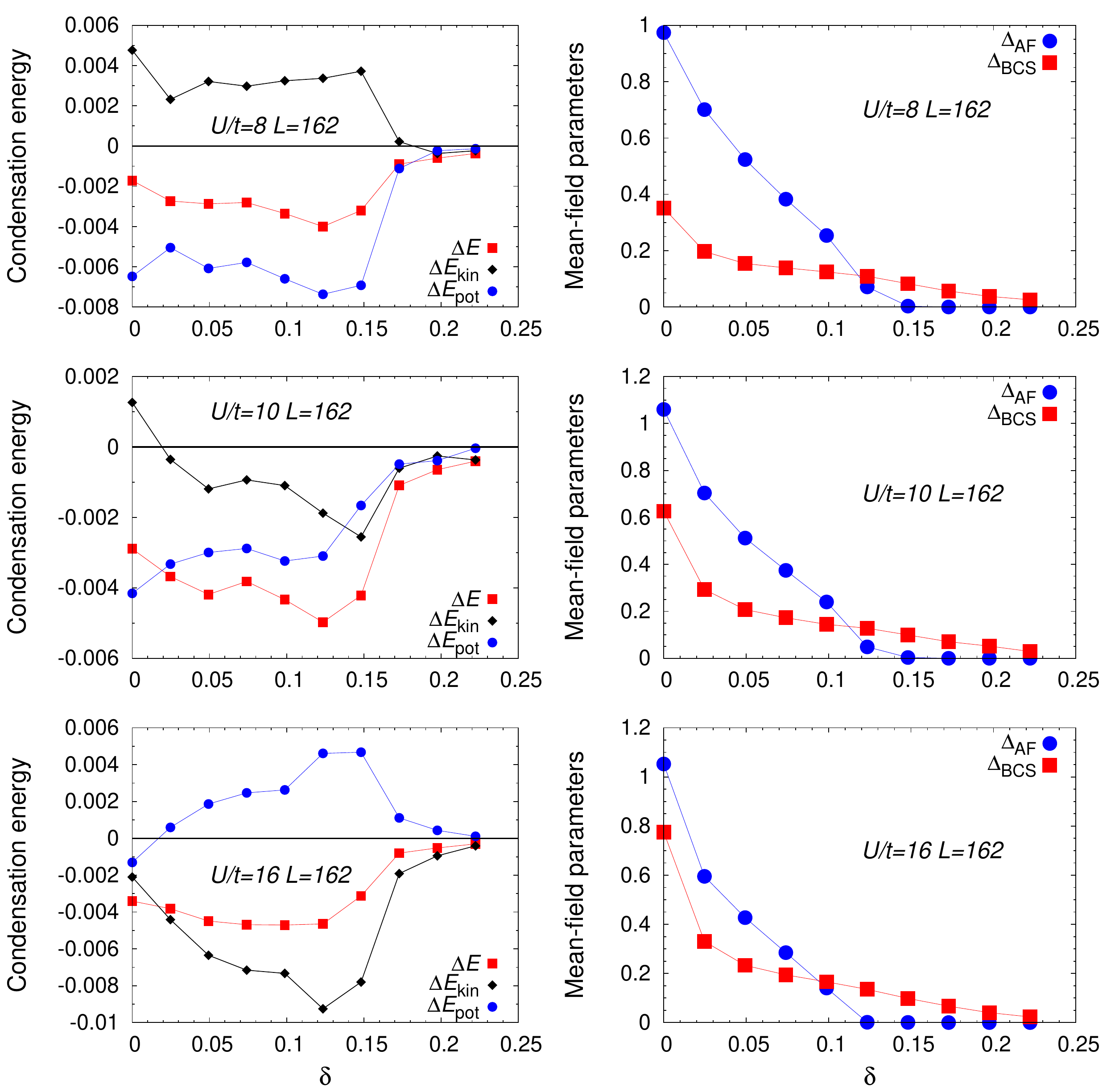}
\caption{\label{fig:Condensation}
(Color online) Left panels: Condensation energy $\Delta E$ (red squares) and its kinetic $\Delta E_{\textrm{kin}}$ (black diamonds) and potential 
$\Delta E_{\textrm{pot}}$ (blue circles) contributions, as a function of doping $\delta$, for $U/t=8$, $10$, and $16$, from top to bottom. 
Right panels: Optimal values of the antiferromagnetic field $\Delta_{\textrm{AF}}$ and the BCS pairing $\Delta_{\textrm{BCS}}$, as a function of 
doping $\delta$, for the same values of $U/t$ that are shown in the left panels. Data are presented on the $L=162$ lattice size.}
\end{figure*}

\begin{figure}
\includegraphics[width=0.8\columnwidth]{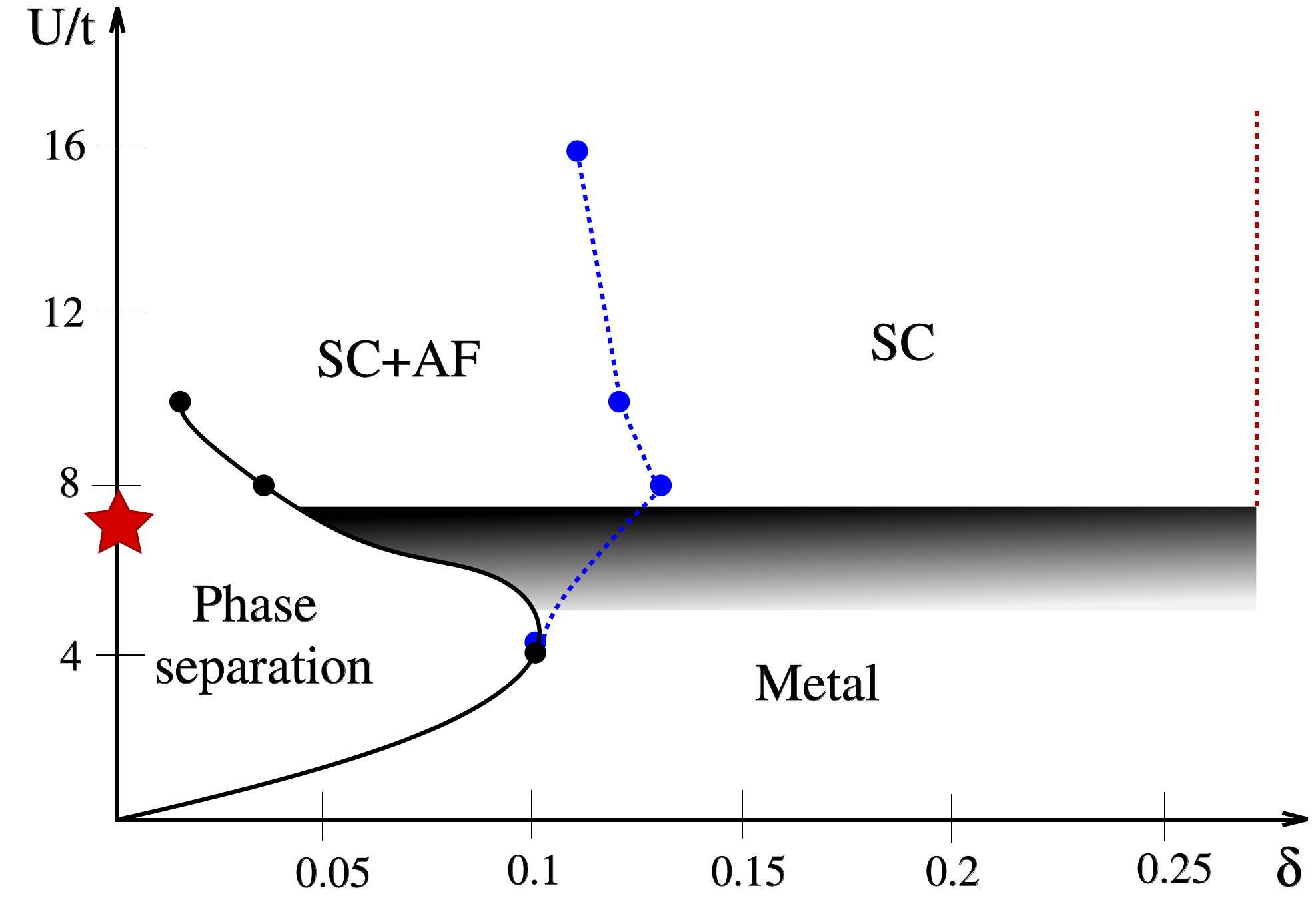}
\caption{\label{fig:pdfinal}
(Color online) Schematic phase diagram as obtained by using a combined VMC and GFMC (with FN approximation) approach. 
The red star labels the location of the hidden Mott transition $U_{\textrm{Mott}}/t$ at half filling. 
The black line with black dots denotes the boundary of the phase-separation region, that shrinks for $U/t \gtrsim U_{\textrm{Mott}}/t$. 
The curve is left open for $U/t > 10$, since we cannot exclude the presence of phase separation very close to half filling. 
The dashed blue line with blue dots marks the disappearance of $\Delta_{\textrm{AF}}$ in the optimal variational state. The dashed red line 
indicates the boundary of the region where sizable pairing correlations are detected. Finally, in the shaded gray region finite-size effects 
are strong and precise results cannot be obtained in the thermodynamic limit.} 
\end{figure}

In order to further analyze the superconducting properties, we consider the condensation energy $\Delta E$, i.e., the energy gain due to the 
inclusion of BCS pairing in the variational state:
\begin{equation}
\Delta E = \frac{\langle\Psi|{\cal H}|\Psi\rangle}{\langle\Psi|\Psi\rangle}-
\frac{\langle\Psi_{\Delta_{\textrm{BCS}}=0}|{\cal H}|\Psi_{\Delta_{\textrm{BCS}}=0}\rangle}
{\langle\Psi_{\Delta_{\textrm{BCS}}=0}|\Psi_{\Delta_{\textrm{BCS}}=0}\rangle},
\end{equation}
where $\Psi_{\Delta_{\textrm{BCS}}=0}$ denotes the best variational state without the inclusion of BCS pairing (but still optimizing 
$\Delta_{\textrm{AF}}$). The results for $\Delta E$ are shown in Fig.~\ref{fig:Condensation} for three values of $U>U_{\rm Mott}$, together 
with the optimal variational parameters $\Delta_{\textrm{BCS}}$ and $\Delta_{\textrm{AF}}$ in $|\Psi\rangle$. We observe that the maximal 
energy gain is obtained close to the point where $\Delta_{\textrm{AF}}$ vanishes. A similar behavior for the antiferromagnetic 
order parameter has been reported also by cellular dynamical mean-field theory~\cite{kancharla2008} and by VMC~\cite{sato2016}. Then, we turn to consider 
the kinetic and potential contributions to the total condensation energy $\Delta E$. For moderate values of $U/t$, i.e., $U/t=8$, the energy 
gain originates from the potential part, while there is a loss in the kinetic part. This feature is consistent with standard BCS theory. 
By increasing the value of the Coulomb repulsion to $U/t=10$, we observe a simultaneous gain in both components of the energy, even though the 
kinetic contribution is smaller than the potential one. Finally, for $U/t=16$, the energy gain is purely kinetic, with a simultaneous loss of 
potential energy (except at half filling), as expected in the strong-coupling limit~\cite{scalapino1998}. A similar behavior of the condensation 
energy as a function of $U$ has been already reported, for instance, in a VMC study of the non-magnetic sector of the Hubbard 
model~\cite{yokoyama2013} and in a diagrammatic expansion of the Gutzwiller wave function~\cite{kaczmarczyk2013}. A slightly different behavior, 
highlighting the existence of a critical doping, has been instead reported in Refs.~\onlinecite{gull2012,fratino2016}. Here, a change in the 
behavior of the condensation energy from kinetic driven to potential driven is observed at a critical value of the doping for intermediate 
values of $U/t$. Nevertheless, all these works indicate that a critical value of the Coulomb repulsion is necessary to observe sizable pairing 
correlations and that, in some region of the phase diagram, superconductivity is kinetic-energy driven, as experimentally suggested by optical 
measurements for underdoped cuprates~\cite{deutscher2005,giannetti2011}.

\section{Conclusions}\label{sec:conc}

In conclusion, our variational approach suggests that the Mott transition, which exists in the paramagnetic sector for 
$\delta=0$~\cite{brinkman1970,georges1996}, may leave an important mark in the more realistic phase diagram, obtained when allowing 
antiferromagnetic long-range order. First of all, our results suggest that a reminiscence of the Mott transition at $U_{\rm Mott}$, hidden by 
the antiferromagnetic phase at half-filling, emerges after a careful analysis of the BCS pairing. This hidden Mott transition is intimately 
related with the change from Slater to Mott antiferromagnetism, the former one being related to a Fermi surface instability towards 
antiferromagnetic order (with a potential energy gain), while the latter one being connected to a super-exchange mechanism (with a kinetic 
energy gain). Most importantly, the Mott antiferromagnet contains electron pairing, as originally suggested by Anderson in the RVB theory of 
superconductivity~\cite{anderson1987}. Within our calculations, it is not clear whether the ``critical'' behavior observed at $U_{\textrm{Mott}}$ 
represents a genuine phase transition, characterized by a thermodynamic (or topological) signal, or it is just a sharp crossover between two 
physically different regimes. Nevertheless, the presence of $U_{\rm Mott}$ has a clear manifestation when doping the system with holes. Indeed, 
for Coulomb interactions that are smaller than this ``critical'' value the system is unstable towards phase separation and there is no strong 
evidence that superconductivity may emerge, even if we cannot exclude the presence of infinitesimal pairing correlations; by contrast, for 
$U>U_{\rm Mott}$, hole doping drives the Mott antiferromagnet into a homogeneous superconducting phase, with the condensation energy gain 
shifting from potential to kinetic by increasing $U/t$. In Fig.~\ref{fig:pdfinal}, we report a schematic phase diagram in the $(\delta,U)$ 
plane. We remark that the presence of the hidden transition (marked by a star) influences the whole phase diagram of the Hubbard model. First 
of all, strong superconducting correlations are present when doping the Mott insulator, which is characterized by the existence of preformed 
electron pairs; in this sense, the picture is similar to the RVB theory, where superconductivity emerge when doping a spin 
liquid~\cite{anderson1987} (here, the only difference is that antiferromagnetic order may coexist with electron pairing). The second effect 
of the ``critical'' point at half filling is to separate a region where phase separation is clearly present from another where it is strongly 
reduced and limited to very small dopings: In this regard, we cannot exclude that phase separation is present for all values of $U/t$, even 
though this must be limited to very small values of dopings close to half filling (for this reason, we do not show the continuation of the 
continuous black line for $U/t>10$). We finally mention that no charge-density waves or stripes have been detected in the cases that have been 
analyzed here (e.g., no strong signals in the density-density correlations have been seen); however, future investigations will consider in 
more detail the possibility that a non-uniform distribution of densities in the variational wave function may lower the VMC or FN energies.

\acknowledgments

We thank M. Fabrizio and F. Mila for useful discussions.

\end{document}